 \documentclass[sort&compress,preprint,review,12pt]{elsarticle}

 \usepackage{graphicx}

\usepackage{amssymb}

\usepackage[usenames]{color}
\journal{Physics Letters A}

\begin{document}

\begin{frontmatter}



\title{Type-I Intermittency With Noise \\ Versus Eyelet Intermittency}


\author[SSU]{Alexander~E.~Hramov}
\author[SSU]{Alexey~A.~Koronovskii}
\author[SSU]{Maria~K.~Kurovskaya}
\author[SSU]{Olga~I.~Moskalenko\corref{lulu}}
\cortext[lulu]{Corresponding Author} \ead{moskalenko@nonlin.sgu.ru}
\address[SSU]{Faculty of Nonlinear
Processes, Saratov State University, 83, Astrakhanskaya, Saratov,
410012, Russia}

\begin{abstract}
In this article we compare the characteristics of two types
of the intermittent behavior (type-I intermittency in the presence
of noise and eyelet intermittency taking place in the vicinity of the chaotic phase synchronization boundary) supposed hitherto to be
different phenomena. We show that these effects are the same type
of dynamics observed under different conditions. The correctness of
our conclusion is confirmed by the consideration of different sample
systems, such as  quadratic map, Van der Pol oscillator and
R\"ossler system. Consideration of the problem concerning the upper boundary of the intermittent behavior also confirms the validity of the statement on the
equivalence of type-I intermittency in the presence of noise and eyelet intermittency observed in the onset of phase synchronization.
\end{abstract}


\end{frontmatter}

\section*{Introduction}
\label{sct:Introduction}

Intermittency is well-known to be an ubiquitous phenomenon in
nonlinear science. Its arousal and main statistical properties have
been studied and characterized already since long time ago, and
different types of intermittency have been classified as types
I--III intermittencies~\cite{Berge:1988_OrderInChaos,Dubois:1983_IntermittencyIII},
on--off
intermittency~\cite{Platt:1993_intermittency,
Hramov:2005_IGS_EuroPhysicsLetters},
eyelet
intermittency~\cite{Pikovsky:1997_EyeletIntermitt,Lee:1998:PhaseJumps,%
Boccaletti:2002_LaserPSTransition_PRL} and ring
intermittency~\cite{Hramov:RingIntermittency_PRL_2006}.

Despite of some similarity (the presence of two different regimes
alternating suddenly with each other in the time series), every type
of intermittency is governed by its own certain mechanisms and
characteristics of the intermittent behavior (such as the dependence
of the mean length of the laminar phases on the control parameter,
the distribution of the laminar phase lengths, etc.) of
different intermittency types are distinct. There are no doubts that
different types of intermittent behavior may take place in a wide
spectrum of systems, including cases of practical interest for
applications in radio engineering, medical, physiological, and other
applied sciences.

This article is devoted to the comparison of characteristics of
type-I intermittency in the presence of noise and eyelet
intermittency taking place in the vicinity of the phase
synchronization boundary. Although these types of
intermittency are known to be characterized by different
theoretical laws, we show here for the first time that
these two types of the intermittent behavior considered hitherto as
different phenomena are, in fact, the same type of the system
dynamics.

The structure of the paper is the following. In
Sec.~\ref{sct:Relation} we give the brief theoretical data
concerning both the type-I intermittency with noise and eyelet
intermittency observed in the vicinity of the phase synchronization boundary as well as arguments confirming the equivalence of these
types of the dynamics. The next Sections~\ref{sct:NumericalVerification}--\ref{sct:VdPunderRsslr} aim to verify the statement given
in the Sec.~\ref{sct:Relation} by means of numerical simulations
of the dynamics of several model systems such as a quadratic map,
R\"ossler oscillators, etc. Eventually, in
Sec.~\ref{sct:Boundary} we discuss the problem of the upper boundary
of the intermittent behavior. The final conclusions are given in
Sec.~\ref{sct:Conclusions}.

\section{Relation between type-I intermittency with noise and eyelet intermittency}
\label{sct:Relation} First, we consider briefly both eyelet
intermittency in the vicinity of the phase
synchronization boundary and type-I intermittency in the presence of noise
following conceptions accepted generally. The main arguments
confirming equivalence of these types of the intermittent behavior
are given afterwards.

\subsection{Type-I intermittency with noise}

The intermittent behavior of type-I is known to be observed below
the saddle-node bifurcation point, with the mean length of laminar
phases $T$ being inversely proportional to the square root of the
criticality parameter $(\varepsilon_c-\varepsilon)$, i.e.
\begin{equation}\label{eq:Type-IIntermittencyPowerLaw}
T\sim(\varepsilon_c-\varepsilon)^{-1/2},
\end{equation}
where $\varepsilon$ is the control parameter and $\varepsilon_c$ is
its bifurcation value corresponding to the bifurcation
point~\cite{Pomeau:1980}. The influence of noise on the system results in the transformation of
characteristics of intermittency~\cite{Eckmann:1981_IntermittencyWithNoise,%
Kye:2000_TypeIAndNoise,Hramov:2007_TypeIAndNoise}, with the
intermittent behavior being observed in this case both below and
above the saddle-node bifurcation point $\varepsilon_c$. In the
supercritical region~\cite{Hramov:2007_TypeIAndNoise} of the control parameter values (i.e., above the
point of bifurcation, ${\varepsilon>\varepsilon_c}$) the mean length
$T$ of the laminar phases is given by
\begin{equation}\label{eq:TypeILawMeanLength}
T=\frac{1}{k\sqrt{\varepsilon-\varepsilon_c}}\exp\left(\frac{4(\varepsilon-\varepsilon_c)^{3/2}}{3D}\right),
\end{equation}
where $k=\mathrm{const}$, $D$ is the intensity of a delta-correlated white noise $\xi_n$
[${\langle\xi_n\rangle=0}$,
${\langle\xi_n\xi_m\rangle=D\delta(n-m)}$],
with Equation~(\ref{eq:TypeILawMeanLength}) being applicable in the region
\begin{equation}
D^{2/3}\ll |\varepsilon-\varepsilon_c|\ll 1
\end{equation}
of the control parameter
plane~\cite{Eckmann:1981_IntermittencyWithNoise,Hramov:ZeroLE_PRE2008}.
In this region the criticality parameter
${(\varepsilon-\varepsilon_c)}$ is large enough and, therefore, the
approximate equation
\begin{equation}\label{eq:TypeILawMeanLengthApprox}
\ln T= B(\varepsilon-\varepsilon_c)^{3/2}-\ln k
\end{equation}
(where $B=\mathrm{const}$) is used typically (see~\cite{Kye:2000_TypeIAndNoise} for detail) instead of~(\ref{eq:TypeILawMeanLength}).
In turn, the distribution $p(\tau)$ of the laminar phase lengths
$\tau$ is governed by the exponential law~\cite{Hramov:2007_TypeIAndNoise}
\begin{equation}\label{eq:LamPahseLengthDistribution}
p(\tau)=T^{-1}\exp\left(-{\tau/T}\right).
\end{equation}

\subsection{Eyelet intermittency}
For the chaotic systems in
the vicinity of the phase synchronization boundary (if the natural
frequencies of oscillator and external signal are detuned slightly)
two types of the intermittent behavior and, correspondingly, two
critical values are reported to
exist~\cite{Pikovsky:1997_EyeletIntermitt,Rosa:1998_TransToPS,Lee:1998:PhaseJumps}.
Below the boundary of the phase synchronization regime the dynamics
of the phase difference $\Delta\varphi(t)$ features time intervals
of the phase synchronized motion (laminar phases) persistently and
intermittently interrupted by sudden phase slips (turbulent phases)
during which the value of $|\Delta\varphi(t)|$ jumps up by $2\pi$.
For two coupled chaotic systems there are two values of the coupling
strength $\varepsilon_1 <\varepsilon_2$ being the characteristic points which
are considered to separate the different types of the dynamics. Below
the coupling strength value $\varepsilon_1$ the type-I intermittency is
observed, with the power law $T\sim(\varepsilon_1-\varepsilon)^{-1/2}$ taking
 place for the mean length of the laminar phases, whereas above
the critical point $\varepsilon_2$ the phase synchronization regime is
revealed. For the coupling strength $\varepsilon\in(\varepsilon_1;\varepsilon_2)$
the super-long laminar behavior (the so called \emph{``eyelet
intermittency''}) should be detected. For eyelet intermittency (see,
e.g.~\cite{Pikovsky:1997_EyeletIntermitt,Lee:1998:PhaseJumps}) the
dependence of the mean length $T$ of the laminar phases on the
criticality parameter is expected to follow the law
\begin{equation}
T\sim\exp\left(\kappa(\varepsilon_2-\varepsilon)^{-1/2}\right)
\end{equation}
or
\begin{equation}\label{eq:EyeletLawMeanLength}
\ln(1/T)=c_0-c_1(\varepsilon_2-\varepsilon)^{-1/2},
\end{equation}
($c_0$, $c_1$ and $\kappa$ are the constants) given for the first time
in~\cite{Grebogi:1983_Basins} for the transient statistics near the
unstable-unstable pair bifurcation point. The analytical form of the
distribution of the laminar phase lengths has not been reported
anywhere hitherto for eyelet intermittency.

The theoretical explanation of the eyelet intermittency phenomenon
is based on the boundary crisis of the synchronous attractor caused
by the unstable-unstable bifurcation when the saddle periodic orbit
and repeller periodic orbit join and
disappear~\cite{Pikovsky:1997_EyeletIntermitt,Rosa:1998_TransToPS}.
This type of the intermittent behavior has been observed both in the
numerical
calculations~\cite{Pikovsky:1997_EyeletIntermitt,Lee:1998:PhaseJumps}
and experimental
studies~\cite{Boccaletti:2002_LaserPSTransition_PRL} for the
different nonlinear systems, including R\"ossler oscillators.

\subsection{Theory of equivalence of the considered types of behavior}
Although type-I intermittency with noise and eyelet intermittency taking place in the vicinity of the chaotic phase synchronization onset
seem to be different phenomena, they are really the same type of the
dynamics observed under different conditions. The difference between these types of the intermittent behavior is
only in the character of the external signal. In case of the type-I
intermittency with noise the stochastic signal influences on the system, while
in the case of eyelet intermittency the signal of chaotic dynamical
system is used to drive the response chaotic oscillator.
At the same time, the core mechanism governed the system behavior (the motion in
the vicinity of the bifurcation point disturbed by the stochastic or
deterministic perturbations) is the same in both cases.
To emphasize the weak difference in the character of the driving signal we shall further use the terms ``type-I intermittency with noise'' and ``eyelet intermittency'' despite of the fact of the equivalence of these types of the intermittent behavior.

Indeed, the phenomena observed near the synchronization boundary for periodic systems whose motion is perturbed by noise (in other words, the behavior in the vicinity of the saddle-node bifurcation perturbed by noise) have been shown recently to be the same as for chaotic oscillators in the vicinity of the phase synchronization boundary~\cite{Hramov:2007_2TypesPSDestruction,Hramov:ZeroLE_PRE2008,Pikovsky:1997_EyeletIntermitt,Hramov:2007_TypeIAndNoise}.
Thus, both for two coupled chaotic R\"ossler systems
and driven Van der Pol oscillator the same scenarios of the
synchronous regime destruction have been
revealed~\cite{Hramov:2007_2TypesPSDestruction}. Moreover, for
two coupled R\"ossler systems the behavior of the conditional
Lyapunov exponent in the vicinity of the onset of the phase
synchronization regime is governed by the same laws as in the case
of the driven Van der Pol oscillator in the presence of
noise~\cite{Hramov:ZeroLE_PRE2008}.
Additionally, when the turbulent phase begins the phase trajectory
demonstrates motion being close to periodic both for the eyelet intermittency observed in the vicinity of the phase synchronization boundary (see~\cite{Rosa:1998_TransToPS}) and for type-I intermittency with noise. Finally, the repeller and saddle periodic orbits of the same period in the vicinity of the parameter region corresponding to the intermittent behavior tend to coalesce with each other (see, e.g.~\cite{Pikovsky:1997_EyeletIntermitt,Hramov:2007_TypeIAndNoise}) for both these types of the intermittent behavior. Obviously, if the phenomena observed near the saddle-node bifurcation point for the systems whose motion is perturbed by noise  are the same as for chaotic oscillators in the vicinity of the phase synchronization onset, one can expect that the intermittent behavior of two coupled chaotic oscillators near the phase synchronization boundary (eyelet intermittency) is also exactly the same as intermittency of type-I in the presence of noise in the supercritical region.

So, if type-I intermittency with noise and eyelet intermittency taking place in the vicinity of the chaotic phase synchronization onset are the same type of the system
dynamics, the theoretical equations~(\ref{eq:TypeILawMeanLengthApprox}) and (\ref{eq:EyeletLawMeanLength}) obtained for these types of the intermittent behavior are the approximate expressions being the different forms of Eq.~(\ref{eq:TypeILawMeanLength}) describing the dependence of the  mean length of the laminar phases on the criticality parameter. Therefore, Eq.~(\ref{eq:EyeletLawMeanLength}) can be deduced from Eq.~(\ref{eq:TypeILawMeanLengthApprox}) and vice versa.
As a consequence, the coefficients $B$, $k$ and $c_0$, $c_1$ in (\ref{eq:TypeILawMeanLengthApprox}) and (\ref{eq:EyeletLawMeanLength}) are related with each other.
Obviously, the mean length of the laminar phases must obey
Eq.~(\ref{eq:TypeILawMeanLengthApprox}) and Eq.~(\ref{eq:EyeletLawMeanLength}) simultaneously, independently whether the system behavior is classified as eyelet intermittency or
type-I intermittency with noise. Additionally, the laminar phase length distribution for the considered type of behavior must satisfy the exponential law~(\ref{eq:LamPahseLengthDistribution}).

The intermittent behavior under study is considered in the coupling strength range ${\varepsilon_c<\varepsilon<\varepsilon_2}$. In the case of the system driven by external noise (type-I intermittency with noise) the lower boundary value $\varepsilon_c$ corresponds to the saddle-node bifurcation point when external noise is switched off. For the dynamical systems demonstrating the chaotic behavior (eyelet intermittency) the lower boundary $\varepsilon_c$ may be found, e.g., in the way described in~\cite{Hramov:ZeroLE_PRE2008}. As far as the choice of the upper boundary value $\varepsilon_2$ is concerned, this subject is discussed in detail in Sec.~\ref{sct:Boundary} of this paper both for chaotic and stochastic external signals.

To find the relationship between coefficients in~(\ref{eq:TypeILawMeanLengthApprox}) and (\ref{eq:EyeletLawMeanLength}) we introduce the auxiliary variable $\xi=(\varepsilon_2-\varepsilon)^{-1/2}$ and expand $\ln(1/T)$ determined by Eq.~(\ref{eq:TypeILawMeanLengthApprox}) (type-I intermittency with noise) into Taylor series in the vicinity of the point $\xi_0=(\varepsilon_2-\varepsilon_0)^{-1/2}$, where ${\varepsilon_c<\varepsilon_0<\varepsilon_2}$, i.e.
\begin{equation}\label{eq:TeylorSeries}
\begin{array}{ll}
\displaystyle\ln\left(1/T\right)&=\left(\ln k-B(\varepsilon_0-\varepsilon_c)^{3/2}\right)-\\
\\
& -3\left(B\sqrt{\varepsilon_0-\varepsilon_c}\left(\varepsilon_2-\varepsilon_0\right)^{3/2}\right)(\xi-\xi_0)+\\
\\
&\displaystyle+\frac{3B(\varepsilon_2-\varepsilon_0)^2(4\varepsilon_0-\varepsilon_2-3\varepsilon_c)}{2\sqrt{\varepsilon_0-\varepsilon_c}}(\xi-\xi_0)^2+o\left((\xi-\xi_0)^3\right)=\\
\\
&\displaystyle=\ln k-\frac{B\left(20\varepsilon_0^2-21\varepsilon_0\varepsilon_2+3\varepsilon_2^2-19\varepsilon_c\varepsilon_0+15\varepsilon_c\varepsilon_2+2\varepsilon_c^2\right)}{2\sqrt{\varepsilon_0-\varepsilon_c}}\\
\\
&\displaystyle -\frac{3B\left(\varepsilon_2-\varepsilon_0\right)^2(5\varepsilon_0-\varepsilon_2-4\varepsilon_c)}{\sqrt{\varepsilon_0-\varepsilon_c}\sqrt{\varepsilon_2-\varepsilon_0}}(\varepsilon_2-\varepsilon)^{-1/2}+\\
\\
&\displaystyle +
\frac{3B\left(\varepsilon_2-\varepsilon_0\right)^2(4\varepsilon_0-\varepsilon_2-3\varepsilon_c)}{2\sqrt{\varepsilon_0-\varepsilon_c}}(\varepsilon_2-\varepsilon)^{-1}+o\left((\xi-\xi_0)^3\right).
\end{array}
\end{equation}
Having neglected the term $o\left((\xi-\xi_0)^3\right)$ in~(\ref{eq:TeylorSeries}) one can write Eq.~(\ref{eq:TeylorSeries}) in the form
\begin{equation}\label{eq:TypeIModified}
\ln(1/T)=c_0-c_1(\varepsilon_2-\varepsilon)^{-1/2}+c_2(\varepsilon_2-\varepsilon)^{-1}.
\end{equation}
Having required
\begin{equation}
c_2=\frac{3B\left(\varepsilon_2-\varepsilon_0\right)^2(4\varepsilon_0-\varepsilon_2-3\varepsilon_c)}{2\sqrt{\varepsilon_0-\varepsilon_c}}\equiv 0
\end{equation}
we obtain, that
\begin{equation}\label{eq:C2coefficientValue}
\varepsilon_0=\frac{\varepsilon_2+3\varepsilon_c}{4}, \qquad \varepsilon_c<\varepsilon_0<\varepsilon_2
\end{equation}
and, therefore, equation~(\ref{eq:TypeIModified}) describing the
dependence of the mean length $T$ of the laminar phases on the
criticality parameter for type-I intermittency with noise coincides exactly with Eq.~(\ref{eq:EyeletLawMeanLength}) corresponding to eyelet intermittency. Correspondingly, in terms of Eq.~(\ref{eq:C2coefficientValue}) the relationship between coefficients $B$, $k$ and $c_0$, $c_1$ in~(\ref{eq:TypeILawMeanLengthApprox}) and (\ref{eq:EyeletLawMeanLength}) is the following
\begin{equation}\label{eq:Coefficients}
\begin{array}{ll}
c_0&=\ln k+B(\varepsilon_2-\varepsilon_c)^{3/2},\\
\\
\displaystyle c_1&=\frac{\displaystyle 9\sqrt{3}}{\displaystyle 16}B(\varepsilon_2-\varepsilon_c)^{2}.
\end{array}
\end{equation}

\begin{figure}[tb]
\centerline{\includegraphics*[scale=0.35]{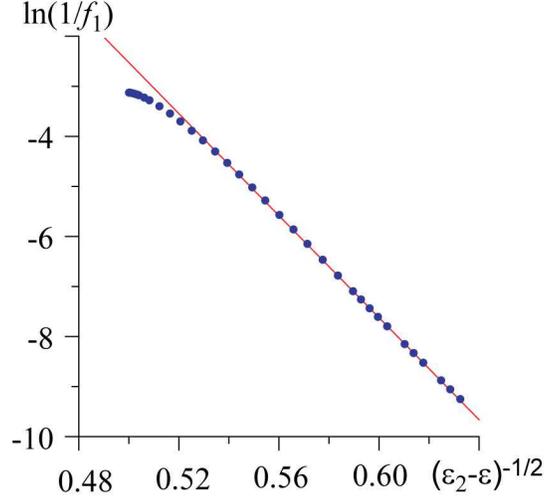}} \caption{The
dependence ${f_1(\varepsilon)=k^{-1}\exp(B \varepsilon^{3/2})}$ simulating the
theoretical law~(\ref{eq:TypeILawMeanLengthApprox}) and its
approximation by the curve $f_2(x)=c_0-c_1(\varepsilon-\varepsilon_2)^{-1/2}$
corresponding to law~(\ref{eq:EyeletLawMeanLength}) for eyelet
intermittency. The parameter values are $k^{-1}=23$, $B=3.33$,
$c_0=23.5$, $c_1=51.9$, $x_2=4.0$. The points corresponding to
$f_1(x)$ are shown by symbols \textcolor{blue}{$\bullet$}. The
theoretical law~(\ref{eq:EyeletLawMeanLength}) for eyelet
intermittency is shown by the solid line} \label{fgr:TwoLaws}
\end{figure}

Fig.~\ref{fgr:TwoLaws} illustrates the relationship of two theoretical laws~(\ref{eq:TypeILawMeanLengthApprox}) and (\ref{eq:EyeletLawMeanLength}) in the region $\varepsilon>\varepsilon_c$. Here function $f_1(x)=k^{-1}\exp(B \varepsilon^{3/2})$ simulates the theoretical law~(\ref{eq:TypeILawMeanLengthApprox}) (the critical point is supposed to be $\varepsilon_c=0$), whereas the curve $f_2(x)=c_0-c_1(\varepsilon-\varepsilon_2)^{-1/2}$ corresponds to law~(\ref{eq:EyeletLawMeanLength}) for eyelet intermittency. The value of coefficients $B$, $k$, $c_0$ and $c_1$ have been selected according to Eq.~(\ref{eq:Coefficients}). One can see that in the region of the study both curves coincide with each other. It means that the mean length of the laminar phases obeys Eq.~(\ref{eq:TypeILawMeanLengthApprox}) and Eq.~(\ref{eq:EyeletLawMeanLength}) simultaneously, independently whether the system behavior is classified as eyelet intermittency or type-I intermittency with noise.


\section{Numerical verifications}
\label{sct:NumericalVerification} To confirm the concept of the
equivalence of intermittencies being the subject of this study we
consider several examples of the intermittent behavior classified
both as eyelet intermittency taking place in the vicinity of the phase synchronization onset (two coupled R\"ossler systems) and
type-I intermittency with noise (quadratic map and driven Van der
Pol oscillator).

\subsection{Two coupled R\"ossler systems}
As we have mentioned above, the intermittent behavior of two coupled
chaotic oscillators in the vicinity of the phase synchronization
boundary is classified traditionally as \emph{eyelet
intermittency}~\cite{Pikovsky:1997_EyeletIntermitt,Rosa:1998_TransToPS,Lee:1998:PhaseJumps}.
Nevertheless, the behavior of two coupled R\"ossler oscillators
close to the phase synchronization onset was considered from
the point of view of type-I intermittency with noise for the first
time in~\cite{Kye:2000_TypeIAndNoise}, whereas the same dynamics
from the position of eyelet intermittency was studied
in~\cite{Lee:1998:PhaseJumps}. According to different works the
mean length of laminar phases happens to satisfy both
Eq.~(\ref{eq:TypeILawMeanLengthApprox})
(Ref.~\cite{Kye:2000_TypeIAndNoise}) and
Eq.~(\ref{eq:EyeletLawMeanLength})
(Ref.~\cite{Lee:1998:PhaseJumps}).
Recently~\cite{Kurovskaya:2008_EyeletVSTypeI_PJTFeng} the
distribution of the laminar phase lengths has been found to obey the
exponential law~(\ref{eq:LamPahseLengthDistribution}) corresponding to type-I intermittency with noise.
To give the
complete picture we replicate the consideration of two coupled
R\"ossler systems near the onset of the phase synchronization regime
for the different type of coupling between oscillators and another
set of the control parameter values and show that the observed
intermittent behavior may be classified both as eyelet intermittency
and type-I intermittency with noise.

The system under study is represented by a pair of unidirectionally
coupled R\"ossler systems, whose equations read as
\begin{equation}
\begin{array}{l}
\dot x_{d}=-\omega_{d}y_{d}-z_{d},\\
\dot y_{d}=\omega_{d}x_{d}+ay_{d},\\
\dot z_{d}=p+z_{d}(x_{d}-c),\\
\\
\dot x_{r}=-\omega_{r}y_{r}-z_{r} +\sigma(x_{d}-x_{r}),\\
\dot y_{r}=\omega_{r}x_{r}+ay_{r},\\
\dot z_{r}=p+z_{r}(x_{r}-c),\\
\end{array}
\label{eq:Roesslers}
\end{equation}
where $(x_{d},y_{d},z_{d})$ [$(x_{r},y_{r},z_{r})$] are the
Cartesian coordinates of the drive [the response] oscillator, dots
stand for temporal derivatives, and $\sigma$ is a parameter ruling
the coupling strength. The other control parameters of
Eq.~(\ref{eq:Roesslers}) have been set to $a=0.15$, $p=0.2$,
$c=10.0$, in analogy with our previous
studies~\cite{Aeh:2005_GS:ModifiedSystem,Harmov:2005_GSOnset_EPL}.
The $\omega_r$--parameter (representing the natural frequency of the
response system) has been selected to be $\omega_r=0.95$; the
analogous parameter for the drive system has been fixed to
$\omega_d=0.93$. For such a choice of the control parameter values,
both chaotic attractors of the drive and response systems are phase
coherent. The instantaneous phase of the chaotic signals
$\varphi(t)$ can be therefore introduced in the traditional way as
the rotation angle ${\varphi_{d,r}=\arctan(y_{d,r}/x_{d,r})}$ on the
projection plane $(x,y)$ of each system.

\begin{figure}[tb]
\centerline{\includegraphics*[scale=0.5]{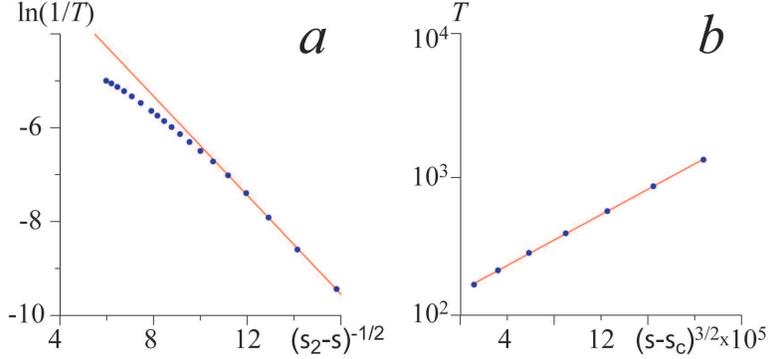}} \caption{The
points obtained numerically for two unidirectionally coupled
R\"ossler oscillators~(\ref{eq:Roesslers}) are shown by symbols
``\textcolor{blue}{$\bullet$}''. The theoretical
laws~(\ref{eq:EyeletLawMeanLength}) and
(\ref{eq:TypeILawMeanLengthApprox}) are shown by the solid lines.
(\textit{a}) \emph{Eyelet intermittency:} the dependence of
$\ln(1/T)$ on the parameter ${(\sigma_2-\sigma)^{-1/2}}$; $\sigma_2=0.042$,
$c_0=-0.75$, $c_1=0.55$. (\textit{b}) \emph{Type-I
intermittency with noise:} the dependence of the mean laminar phase
length $T$ on the parameter ${(\sigma-\sigma_c)^{3/2}}$, with the
ordinate axis being shown in the logarithmic scale; $\sigma_c=0.0345$,
$B=1.01\times10^{4}$, $k=6.68\times 10^{-4}$} \label{fgr:RsslrsGraph}
\end{figure}

In Fig.~\ref{fgr:RsslrsGraph} one and the same result of the
numerical simulation of two coupled R\"ossler
systems~(\ref{eq:Roesslers}) is shown in different ways to compare
obtained data with the analytical predictions~(\ref{eq:EyeletLawMeanLength}) and
(\ref{eq:TypeILawMeanLengthApprox}) for eyelet
intermittency taking place near the phase synchronization boundary (Fig.~\ref{fgr:RsslrsGraph},\,\textit{a}) and type-I
intermittency with noise (Fig.~\ref{fgr:RsslrsGraph},\,\textit{b}), respectively.
The dependence of $T$ on ${(\sigma_2-\sigma)}$ is shown in the whole
range of the coupling parameter strength values
(Fig.~\ref{fgr:RsslrsGraph},\,\textit{a}) to make evident the
deviation of numerically obtained data from
law~(\ref{eq:EyeletLawMeanLength}) far away from the onset of the
phase synchronization. The coupling strength $\sigma$ plays the role
of the control parameter. The
critical point $\sigma_2\approx0.042$ relates to the onset of the
phase synchronization regime in two coupled R\"ossler systems. The
point $\sigma_c\approx0.0345$ used in~(\ref{eq:TypeILawMeanLength})
and (\ref{eq:TypeILawMeanLengthApprox}) corresponds to the
saddle-node bifurcation point if the chaotic dynamics being the
analog of noise could be switched off. The value of this point has
been found from the dependence of the zero conditional Lyapunov
exponent on the coupling strength (see for
detail~\cite{Hramov:ZeroLE_PRE2008}).

One can see, that the intermittent behavior of two coupled R\"ossler
systems may be treated both as eyelet and noised type-I
intermittency with the excellent agreement between numerical data
and theoretical curve in both cases. Moreover, the coefficients $c_0$, $c_1$ and $B$, $k$ of the theoretical equations~(\ref{eq:EyeletLawMeanLength}) and
(\ref{eq:TypeILawMeanLengthApprox}) agree very well with each other according to Eq.~(\ref{eq:Coefficients}).
It allows us to state that both
these effects are the same type of the system dynamics.
Nevertheless, to be totaly convinced of the correctness of our
decision we have to consider other examples of the intermittent
behavior classified traditionally (contrary to the previous case of
two coupled R\"ossler systems) as type-I intermittency with noise.

\subsection{Driven Van der Pol oscillator with noise}
The second sample dynamical system to be considered is Van der Pol
oscillator
\begin{equation}
{\ddot{x}-(\lambda-x^2)\dot{x}+x=A\sin(\omega_et)+D\xi(t)}
\label{eq:DrivenVdPOscillatorAndNoise}
\end{equation}
driven by the external harmonic signal with the amplitude $A$ and
frequency $\omega_e$ with the added stochastic term $D\xi(t)$. The
values of the control parameters have been selected as
$\lambda=0.1$, $\omega_e=0.98$. For the selected values of the
control parameters and $D=0$ the dynamics of the driven Van der Pol
oscillator becomes synchronized when ${A=A_c=0.0238}$ that
corresponds to the saddle-node bifurcation. The probability density
of the random variable $\xi(t)$ is
\begin{equation}
p(\xi)=\frac{1}{\sqrt{2\pi}\sigma}\exp\left(-\frac{\xi^2}{2\sigma^2}\right),
\label{eq:NormalDistrib}
\end{equation}
where $\sigma^2=1$. To integrate
Eq.~(\ref{eq:DrivenVdPOscillatorAndNoise}) the one-step Euler method
has been used with time step $h=5\times10^{-4}$, the value of the
noise intensity has been fixed as $D=1$.

\begin{figure}[tb]
\centerline{\includegraphics*[scale=0.5]{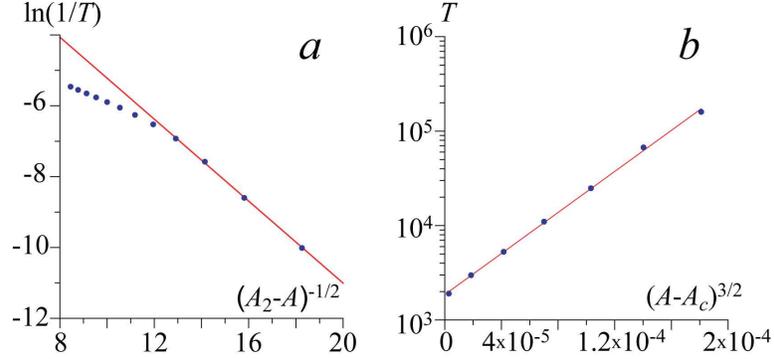}} \caption{The
points obtained numerically for driven Van der Pol oscillator with
the stochastic force~(\ref{eq:DrivenVdPOscillatorAndNoise}) are
shown by symbols ``\textcolor{blue}{$\bullet$}''. The theoretical
laws~(\ref{eq:EyeletLawMeanLength}) and
(\ref{eq:TypeILawMeanLengthApprox}) are shown by the solid lines.
(\textit{a}) \emph{Eyelet intermittency:} the dependence of
$\ln(1/T)$ on the parameter ${(A_2-A)^{-1/2}}$; $A_2=0.029$, $c_0=1.24$, $c_1=0.62$.
(\textit{b}) \emph{Type-I intermittency with noise:}
the dependence of the mean laminar phase length $T$ on the parameter
${(A-A_c)^{3/2}}$; $A_c=0.0238$, $B=2.37\times10^{4}$, $k=4.76\times 10^{-4}$}
\label{fgr:VdPGraph}
\end{figure}

On the one hand, as it has been discussed above, the intermittent
behavior in this case have to be classified as type-I intermittency
with noise. The corresponding dependence of the mean length of
laminar phases on the criticality parameter ${(A-A_c)}$ is shown in
Fig.~\ref{fgr:VdPGraph},\,\textit{b}. If the amplitude $A$ of the
external signal exceeds the critical value $A_c$ the exponential law
${T\sim\exp(\alpha(A-A_c)^{3/2})}$ is expected to be observed. To make
this law evident the abscissa in
Fig.~\ref{fgr:VdPGraph},\,\textit{b} has been selected in the
$(A-A_c)^{3/2}$-scale and the ordinate axis $T$ is shown in the
logarithmic scale. One can see again the excellent agreement between
the numerically calculated data and theoretical
prediction~(\ref{eq:TypeILawMeanLengthApprox}). The distribution of
the lengths of the laminar phases $p(t)$ obtained for $A>A_c$ also
confirms the theoretical
curve~(\ref{eq:LamPahseLengthDistribution}), see Fig.~7
in~\cite{Hramov:2007_TypeIAndNoise}.

On the other hand, trying to choose the corresponding values of
$A_2$ for the driven Van der Pol
oscillator~(\ref{eq:DrivenVdPOscillatorAndNoise}) one can find out
that the intermittent behavior of this system also may be identified
as eyelet intermittency. Indeed, in
Fig.~\ref{fgr:VdPGraph},\,\textit{a} one can see a very good
agreement between the numerically obtained mean length $T$ of the
laminar phases for the different values of the coupling parameter
and theoretical law~(\ref{eq:EyeletLawMeanLength}) corresponding to
the eyelet intermittency. Note also, that for the well chosen values
of $A_2$ the dependence $T(A_2-A)$ in the axes
$((A_2-A)^{-1/2},\ln(1/T))$ behaves in the same way as the
corresponding function $T(\sigma_2-\sigma)$ in the axes
$((\sigma_2-\sigma)^{-1/2},\ln(1/T))$ for two coupled R\"ossler
systems~(\ref{eq:Roesslers}). Again, as well as for two coupled R\"ossler oscillators,
the coefficients $c_0$, $c_1$ and $B$, $k$ of the theoretical equations~(\ref{eq:EyeletLawMeanLength}) and
(\ref{eq:TypeILawMeanLengthApprox}) agree very well with each other according to Eq.~(\ref{eq:Coefficients}).

\subsection{Quadratic map with stochastic force}

The next example is the quadratic map
\begin{equation}
x_{n+1}=x_n^2+\lambda-\varepsilon+D\xi_n, \qquad \mathrm{mod}~1,
\label{eq:QuadraticMap}
\end{equation}
where the operation of ``$\mathrm{mod~}1$'' is used to provide the
return of the system in the vicinity of the point $x=0$,
${\lambda=0.25}$ and the probability density of the stochastic
variable $\xi$ is distributed uniformly throughout the interval
${\xi\in[-1,1]}$. If the intensity of noise $D$ is equal to zero the
saddle-node bifurcation is observed for $\varepsilon=0$. The
intermittent behavior of type-I is observed for $\varepsilon<0$,
whereas the stable fixed point takes place for $\varepsilon>0$.
Having added the stochastic force (${D>0}$)
in~(\ref{eq:QuadraticMap}) we suppose that the intermittent behavior
must be also observed in the area of the positive values of the
criticality parameter $\varepsilon$, with the mean length of the
laminar phases obeying law~(\ref{eq:TypeILawMeanLengthApprox}).

\begin{figure}[tb]
\centerline{\includegraphics*[scale=0.5]{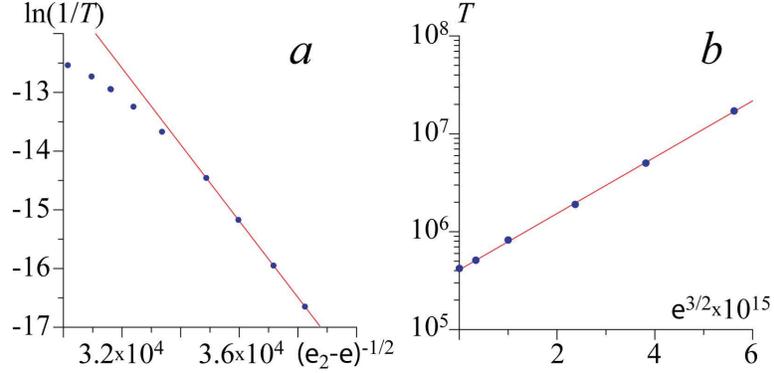}} \caption{The
points obtained numerically for quadratic
map~(\ref{eq:QuadraticMap}) are shown by symbols
``\textcolor{blue}{$\bullet$}''. The theoretical
laws~(\ref{eq:EyeletLawMeanLength}) and
(\ref{eq:TypeILawMeanLengthApprox}) are shown by the solid lines.
(\textit{a}) \emph{Eyelet intermittency:} the dependence of
$\ln(1/T)$ on the parameter ${(\varepsilon_2-\varepsilon)^{-1/2}}$;
$\varepsilon_2=10^{-9}$, $c_0=7.9$, ${c_1=6.45\times10^{-4}}$. (\textit{b})
\emph{Type-I intermittency with noise:} the dependence of the mean
laminar phase length $T$ on the parameter ${\varepsilon^{3/2}}$;
$\varepsilon_c=0$, $D=10^{-7}$, $B=6.62\times10^{14}$, $k=2.35\times 10^{-6}$}
\label{fgr:QMapGraph}
\end{figure}

Although in this case we deal with type-I intermittency with noise,
the numerically obtained points corresponding to the mean length of
laminar phases are approximated successfully both by
Eq.~(\ref{eq:EyeletLawMeanLength}) and
(\ref{eq:TypeILawMeanLengthApprox}) (see Fig.~\ref{fgr:QMapGraph}),
with the coefficients $c_0$, $c_1$ and $B$, $k$ of the theoretical equations~(\ref{eq:EyeletLawMeanLength}) and
(\ref{eq:TypeILawMeanLengthApprox}) agreeing with each other according to Eq.~(\ref{eq:Coefficients}).
These findings confirm our statement about identity of the considered types of
the intermittent behavior.

So, having studied the intermittent behavior of different systems
which (based on the prior knowledge) should be classified either
eyelet intermittency in the vicinity of the phase synchronization boundary or type-I intermittency with the presence of
noise, we can conclude that the obtained characteristics are exactly
the same in all cases described above. Two next sections are devoted
to the consideration of another systems to give the additional
proofs of the correctness of the introduced concept.

\section{Van der Pol oscillator driven by the chaotic signal}\label{sct:VdPunderRsslr}
In this section we consider Van der Pol oscillator driven by the
chaotic signal of R\"ossler system
\begin{equation}\label{eq:VdPUnderRsslr}
\begin{array}{l}
\dot x_{d}=\alpha(-\omega y-z),\\
\dot y_{d}=\alpha(\omega x+ay),\\
\dot z_{d}=\alpha(p+z(x-c)),\\
\ddot{u}-(\lambda-u^2)\dot{u}+u=\varepsilon(Dy-\dot{u}),\\
\end{array}
\end{equation}
where ${a=0.15}$, ${p=0.2}$, ${c=10}$, $\lambda=0.1$,
$\omega=0.9689$ are the control parameters. The auxiliary parameters
$\alpha=0.99$ and $D=0.0664$ alter the characteristics (the
amplitude and main frequency) of the chaotic signal influencing on
Van der Pol oscillator.

From the formal point of view the behavior of
system~(\ref{eq:VdPUnderRsslr}) can be classified neither eyelet intermittency nor
type-I intermittency with noise. Indeed, since the response
oscillator is periodic there are no unstable periodic orbits
embedded into its attractor to be synchronized, therefore, the
system dynamics can not be considered as eyelet intermittency.
Alternatively, due to the presence of chaotic perturbations there is
no pure saddle-node bifurcation in this system to say about type-I
intermittency. Nevertheless, it is intuitively clear that this
example is nearly related to all cases considered above and one can
expect to observe here the same type of intermittency as before.

\begin{figure}[tb]
\centerline{\includegraphics*[scale=0.5]{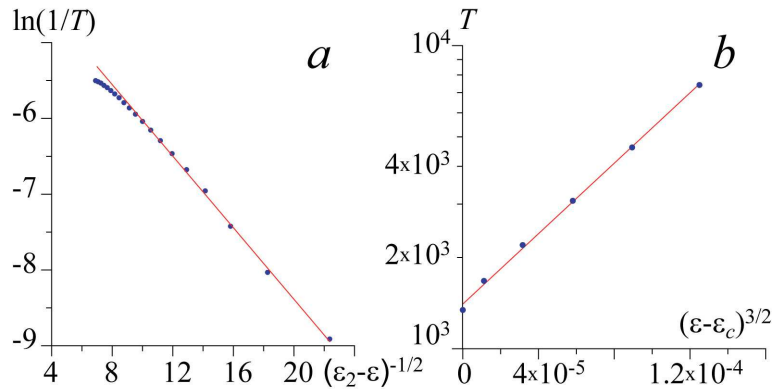}} \caption{The
points obtained numerically for Van der Pol oscillator driven by
chaotic signal of R\"ossler system~(\ref{eq:VdPUnderRsslr}) are
shown by symbols ``\textcolor{blue}{$\bullet$}''. The theoretical
laws~(\ref{eq:EyeletLawMeanLength}) and
(\ref{eq:TypeILawMeanLengthApprox}) are shown by the solid lines.
(\textit{a}) \emph{Eyelet intermittency:} the dependence of
$\ln(1/T)$ on the parameter ${(\varepsilon_2-\varepsilon)^{-1/2}}$;
$\varepsilon_2=0.023$, $c_0=-3.50$, ${c_1=2.46\times10^{-1}}$. (\textit{b})
\emph{Type-I intermittency with noise:} the dependence of the mean
laminar phase length $T$ on the parameter
${(\varepsilon-\varepsilon_c)^{3/2}}$; $\varepsilon_c=0.0185$, $B=1.25\times10^{4}$, $k=6.90\times 10^{-4}$} \label{fgr:VdPUnderRsslrGraph}
\end{figure}

Fig.~\ref{fgr:VdPUnderRsslrGraph} makes this statement evident.
Indeed, the numerically obtained data obey both
laws~(\ref{eq:EyeletLawMeanLength}) and
(\ref{eq:TypeILawMeanLengthApprox}), with
the coefficients $c_0$, $c_1$ and $B$, $k$ of the theoretical equations~(\ref{eq:EyeletLawMeanLength}) and
(\ref{eq:TypeILawMeanLengthApprox}) agreeing with each other in accordance with Eq.~(\ref{eq:Coefficients}).
Additionally, the distribution of the
lengths of the laminar phases follows the exponential
law~\cite{Moskalenko:2007_VdPUnderRsslr_PJTFeng}, that allows us to
say that we deal here with the same type of the dynamics as in the
cases of quadratic map~(\ref{eq:QuadraticMap}), driven Van der Pol
oscillator~(\ref{eq:DrivenVdPOscillatorAndNoise}) and two coupled
R\"ossler systems~(\ref{eq:Roesslers}) considered above.

\section{Upper boundary of the intermittent behavior}\label{sct:Boundary}

All arguments given above may be considered as the evidence of the
proposed statement on the equivalence of both types of the
intermittent behavior. At the same time, one great difference
between type-I intermittency with noise and eyelet intermittency taking place near the onset of the phase synchronization
seems to exist. This difference is connected with the upper boundary
of the intermittent behavior and this point could refute the main
statement of this manuscript. Indeed, for type-I intermittency with
noise in the supercritical region there is no an upper threshold
(see Eq.~(\ref{eq:Type-IIntermittencyPowerLaw})) and the
intermittent behavior may be (theoretically) observed for arbitrary
values of the criticality parameter $(\varepsilon-\varepsilon_c)>0$,
although the length of the laminar phases may be extremely long in
this case, depending on the ratio between the criticality parameter
value and the noise intensity. Alternatively, the existence of the
boundary of the phase synchronization regime being the upper border
of the eyelet intermittency is believed to be undeniable, since there is a
great amount of works where the boundary of the phase
synchronization had been observed and determined. So, this
circumstance along with the arguments given above in
Sections~\ref{sct:Relation}--\ref{sct:NumericalVerification} involve a
seeming contradiction. To resolve this disagreement we consider the probability $P(L,\varepsilon)$ to
observe the turbulent phase in the time realization of the system
demonstrating type-I intermittency with noise in the supercritical
region ${\varepsilon>\varepsilon_c}$ during the observation interval with the length $L$.

The probability to detect the turbulent phase depends on the length $L$ of the observation interval and the length $\tau$ of the laminar phase being realized at the beginning of the system behavior examination. Obviously, if ${\tau<L}$ the turbulent phase is detected with the probability $P_{\tau<L}^{det}=1$ and, in turn, the probability for the laminar phase with length $\tau<L$ to be realized is
\begin{equation}
P(\tau<L)=\int\limits_0^L p(\tau)\,d\tau,
\end{equation}
where $p(\tau)$ is given by Eq.~(\ref{eq:LamPahseLengthDistribution}). Otherwise, when ${\tau>L}$ the probability to detect the turbulent phase is ${P_{\tau>L}^{det}=L/\tau}$, whereas the laminar phase with the length $\tau$ takes place with the probability $P(\tau)=p(\tau)\,d\tau$. Correspondingly, the probability to observe the turbulent phase is
\begin{equation}\label{eq:ProbabilitySurface}
\begin{array}{ll}
\displaystyle P(L,\varepsilon)& \displaystyle =\int\limits_0^L P_{\tau<L}^{det}\, p(\tau)\,d\tau + \int\limits_{L}^{+\infty}P_{\tau>L}^{det} \, p(\tau)\,d\tau=\\
\\
& \displaystyle =\int\limits_0^L \frac{e^{\textstyle-\frac{\tau}{T(\varepsilon)}}}{T(\varepsilon)}\,d\tau+ \int\limits_{L}^{+\infty}\frac{L}{\tau}\frac{e^{\textstyle-{\frac{\tau}{T(\varepsilon)}}}}{T(\varepsilon)}\,d\tau=\\
\\
 & =\displaystyle 1-e^{\textstyle-\frac{L}{T(\varepsilon)}}+\frac{L}{T(\varepsilon)}\times\Gamma\left(0,\frac{L}{T(\varepsilon)}\right),
\end{array}
\end{equation}
where $\Gamma(a,z)$ is the incomplete gamma function, $T(\varepsilon)$ is the mean length
of laminar phases depending on the criticality parameter $\varepsilon$ and given by Eq.~(\ref{eq:TypeILawMeanLength}).

\begin{figure}[p]
\centerline{\includegraphics*[scale=0.45]{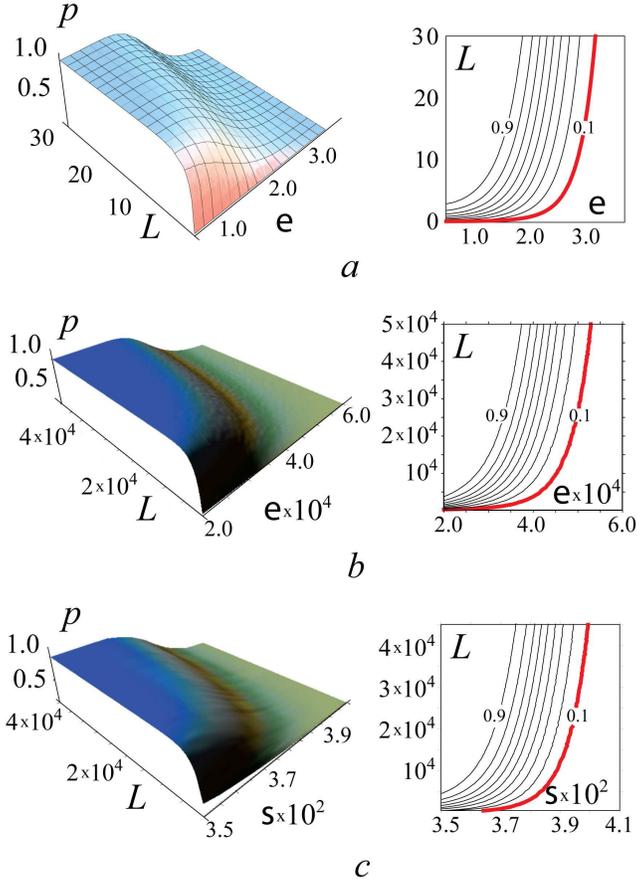}} \caption{The
surface of probability to detect the turbulent phase in the time
realization of the system during the observation interval with the
length $L$ and corresponding to it the level curves with the step
$\Delta P=0.1$. The level curves $P_b=0.1$ demarcating the regions
with the high and low probabilities to detect the turbulent phase
are shown by solid lines. (\textit{a}) The theoretical
expression~(\ref{eq:ProbabilitySurface}), for the simplicity the
values of the control parameters in
Eq.~(\ref{eq:TypeILawMeanLength}) are taken $D=1$, $k=1$,
$\varepsilon_c=0$. (\textit{b}) The probability surface
$P(L,\varepsilon)$ for the circle map with
noise~(\ref{eq:CircleMapSeries}). (\textit{c}) The probability
surface $P(L,\varepsilon)$ for two coupled R\"ossler
systems~(\ref{eq:Roesslers})} \label{fgr:ProbabilitySurfaces}
\end{figure}

The surface $P(L,\varepsilon)$ determined by the analytical
expression~(\ref{eq:ProbabilitySurface}) and level curves corresponding to it
are shown in Fig.~\ref{fgr:ProbabilitySurfaces},\textit{a}. It is clear that the
probability to detect the turbulent phase for
type-I intermittency with noise during one observation grows with the increase of the
examination length $L$ but decreases when the criticality parameter
${\varepsilon-\varepsilon_c}$ is enlarged. Obviously, if one
examines (experimentally or numerically) the system behavior in the
time interval with the length $L$ varying the control parameter
$\varepsilon$, one observes the alternation of the laminar and
turbulent phases for the relatively small values of the
$\varepsilon$-parameter, where $P(L,\varepsilon)$ is close to one,
and only the laminar behavior for the relatively large ones, where
$P(L,\varepsilon)$ is close to zero. Having no information about the
kind of intermittency (e.g., when the experimental study of some
system is carried out) one can suppose the presence of the boundary
separating two different types (intermittent and steady) of dynamics
and, moreover, find a value $\varepsilon_2$ corresponding to the
``onset'' of the laminar behavior. Evidently, this ``boundary
point'' would be correspond to the low probability
$P(L,\varepsilon)$, say, e.g. $P_b=P(L,\varepsilon)=0.1$. In addition,
one can perform ``more careful'' measurements with the increased
length $L'$ of the observation to determine the value of the
boundary point $\varepsilon_2$ more precisely. In this case a new
value $\varepsilon_2(L')$ would be obtained
(${P(L',\varepsilon_2(L'))\approx P_b}$), with it being slightly
larger than the previous one. The schematic location of the ``boundary''
curve $\varepsilon_2(L)$ on the plane $(\varepsilon,L)$ is shown in
Fig.~\ref{fgr:ProbabilitySurfaces},\textit{a} by the solid line. It
is clearly seen, that for the $P_b$-level the length $L$ grows
extremely rapidly with the increase of the $\varepsilon$-value. In
other words, the major extensions of the observation interval $L$
result in the minor corrections of the ``boundary'' point
$\varepsilon_2$. Since the resources of the both experimental and
numerical studies are always limited, some final value
$\varepsilon_2$ with the maximal possible accuracy will be
eventually found. So, despite the fact, that for the type-I
intermittency with noise in the supercritical region the turbulent
phases can always be observed theoretically, from the practical
point of view (in the experimental studies or numerical
calculations) the boundary point $\varepsilon_2$ exists, above which
only the laminar behavior is observed. Moreover, with the further
development of the experimental and computational resources the
additional studies would result only in the insufficient increase of
the boundary value.

To illustrate the drawn conclusion we consider the circle map
\begin{equation}\label{eq:CircleMap}
x_{n+1}=x_n+2\Omega(1-\cos x_n)-\varepsilon+\xi_n, \qquad\mathrm{mod}~2\pi
\end{equation}
in the interval $x\in[-\pi,\pi)$, where $\varepsilon$ is the control
parameter, $\Omega=1.0$, $\xi_n$ is supposed to be a
delta-correlated Gaussian white noise [${\langle\xi_n\rangle=0}$,
${\langle\xi_n\xi_m\rangle=D\delta(n-m)}$]. If the intensity of
noise $D$ is equal to zero, the saddle-node bifurcation is observed
in~(\ref{eq:CircleMap}) for $\varepsilon=\varepsilon_c=0$, when the
stable and unstable fixed points annihilate at $x=0$. Obviously, for
the selected value of the control parameter $\Omega$ the evolution
of system~(\ref{eq:CircleMap}) in the vicinity of the bifurcation
point may be reduced to the quadratic map
\begin{equation}\label{eq:CircleMapSeries}
x_{n+1}=x^2_n+\frac{1}{4}-\varepsilon+\xi_n,
\end{equation}
allowing an easy comparison with the results given in the previous
sections.

The intermittent behavior of type-I is observed for
${\varepsilon<0}$, whereas the stable fixed point takes place for
${\varepsilon>0}$. For the added stochastic force
(${D=4\times10^{-6}}$) in circle map~(\ref{eq:CircleMap}) the
intermittent behavior is also observed in the supercritical region
of the criticality parameter $\varepsilon$, with the mean length $T$
of the laminar phases and the distribution $p(\tau)$ of the laminar
phase lengths $\tau$ obeying
laws~(\ref{eq:TypeILawMeanLengthApprox}) and
(\ref{eq:LamPahseLengthDistribution}), respectively.

The surface of the probability $P(L,\varepsilon)$ to observe the
turbulent phase for the circle map~(\ref{eq:CircleMap}) as well as
the corresponding level curves are shown in
Fig.~\ref{fgr:ProbabilitySurfaces},\textit{b}. To obtain this
surface we have made ${N=2.5\times10^{4}}$ observations for every
point taken with the steps $\Delta L=2\times10^2$,
$\Delta\varepsilon=10^{-5}$ on the parameter plane
$(\varepsilon,L)$. The probability $P(L,\varepsilon)$ was calculated
as ${P(L,\varepsilon)=N_1(L,\varepsilon)/N}$, where
$N_1(L,\varepsilon)$ is the number of observations for which the
turbulent phase has been detected. One can see the excellent
agreement between the results of numerical calculations and
theoretical predictions (compare
Fig.~\ref{fgr:ProbabilitySurfaces},\textit{a} and \textit{b}).

Similarly, the analogous probability surface $P(L,\sigma)$ and the level curves shown in
Fig.~\ref{fgr:ProbabilitySurfaces},\textit{c} have been calculated for
two coupled R\"ossler systems~(\ref{eq:Roesslers}) in the vicinity
of the phase synchronization boundary, where eyelet intermittency is
observed. In this case $N=10^3$ observations have been made for
every point to be examined, with these points being taken with the steps
$\Delta L=10^4$ and $\Delta\sigma=10^{-3}$ on the plane
${(\sigma,L)}$. It is easy to see that for the eyelet intermittency
the probability surface as well as the level curves are exactly the
same as for type-I intermittency with noise in the supercritical
region. As a consequence, we can draw a conclusion, that the eyelet
intermittency taking place in the vicinity of the phase synchronization boundary and type-I intermittency with noise in the
supercritical region are the same type of the dynamics observed under
different conditions. Another consequence of the made consideration
is the fact, that the phase synchronization boundary point can not
be found absolutely exactly, since it separates the regions with the
high and low probabilities to observe the phase slips in the coupled
chaotic systems with the help of the experimental and computational
resources existing at the moment of study. If someone, using a more
powerful tools, tried to refine, say, the value of the coupling
strength corresponding to the phase synchronization boundary
reported in the earlier paper, one would obtain a new value being
close to the previous one, but larger. Exactly the same situation
may be found, e.g., in the work~\cite{Pazo:2002_UPOsSynchro}, where
two mutually coupled R\"ossler systems have been considered. In this work the
refined boundary value $\varepsilon_{PS}=0.0416$ is reported with
the reference to the earlier work~\cite{Rosenblum:1997_LagSynchro},
where the value $\varepsilon_{PS}=0.036$ was given.

\section{Conclusions}
\label{sct:Conclusions}

Having considered two types of the intermittent behavior, namely
eyelet intermittency taking place in the vicinity of the phase synchronization boundary and type-I intermittency with noise, supposed
hitherto to be different, we have shown that these effects are the same
type of the dynamics observed under different conditions. The analytical relation between coefficients of the theoretical equations corresponding to both types of the intermittent behavior has been obtained.

The difference between these types of the intermittent behavior is
only in the character of the external signal. In case of the type-I
intermittency the stochastic signal influences on the system, while
in the case of eyelet intermittency the signal of chaotic dynamical
system is used to drive the response chaotic oscillator\footnote{Since chaotic regime has some memory in contrast to white noise, perhaps, it would be more appropriate to use the colored noise with a comparable memory for type-I intermittency with noise to compare characteristics of the both types of intermittent behavior. At the same time, our studies show that the character of noise (such as distribution, the presence of memory) does not influence sufficiently on the characteristics of intermittency, and, therefore, the simpler model of noise may be used}. At the same
time, the core mechanism governed the system behavior as well as the
characteristics of the system dynamics are the same in both cases.

\section*{Acknowledgments}
we thank the Referees for useful comments and remarks. This work has been supported by Federal special-purpose programme
``Scientific and educational personnel of innovation Russia (2009--2013)''
and the President Program (NSh-3407.2010.2).



\end{document}